\documentclass[longauth, letter]{aa} 

\usepackage{amsmath}
\usepackage{siunitx}

\usepackage{graphicx}
\usepackage{float}
\usepackage{txfonts}

\newcommand{\kms}{km\,s$^{-1}$\xspace}
\newcommand{\msun}{M\ensuremath{_\odot}\xspace}
\newcommand{\jybm}{Jy\,beam$^{-1}$\xspace}

\usepackage{hyperref}
\hypersetup{colorlinks,linkcolor=blue,citecolor=blue,filecolor=cyan,urlcolor=blue}

\begin{document} 

   \titlerunning{Evidence for substructure in a young Class 0 disk}
   \title{FAUST. XVIII. Evidence for annular substructure in a very young Class 0 disk.}

     \authorrunning{Maureira et al.}
     \author{M. J. Maureira\inst{1}, J. E. Pineda \inst{1}, H. B. Liu \inst{2}, L. Testi \inst{3,4}, D. Segura-Cox \inst{5}, C. Chandler \inst{6}, D. Johnstone \inst{7,8}, P. Caselli \inst{1}, G. Sabatini \inst{4}, Y. Aikawa \inst{9},  E. Bianchi \inst{10}, C. Codella \inst{4,11}, N. Cuello \inst{11}, D. Fedele \inst{4}, R. Friesen \inst{12}, L. Loinard \inst{13}, L. Podio \inst{4}, C. Ceccarelli \inst{11}, N. Sakai \inst{12}, 
          \and
         S. Yamamoto \inst{13}
          }

   \institute{Max-Planck-Institut für extraterrestrische Physik (MPE), Gießenbachstr. 1, D-85741 Garching, Germany, 
              \email{maureira@mpe.mpg.de}
         \and Department of Physics, National Sun Yat-Sen University, No. 70, Lien-Hai Road, Kaohsiung City, 80424, Taiwan, ROC ; Center of Astronomy and Gravitation, National Taiwan Normal University, Taipei, 116, Taiwan, ROC
         \and Dipartimento di Fisica e Astronomia “Augusto Righi”, Viale Berti Pichat 6/2, Bologna, Italy
         \and INAF, Osservatorio Astrofisico di Arcetri, Largo E. Fermi 5, 50125, Firenze, Italy
         \and Department of Astronomy, The University of Texas at Austin, 2515
Speedway, Austin, TX 78712, USA
          \and National Radio Astronomy Observatory, PO Box O, Socorro, NM 87801, USA
          \and NRC Herzberg Astronomy and Astrophysics, 5071 West Saanich Road, Victoria, BC V9E 2E7, Canada
          \and Department of Physics and Astronomy, University of Victoria, Victoria, BC V8P 5C2, Canada
          \and Department of Astronomy, The University of Tokyo, 7-3-1 Hongo, Bunkyo-ku, Tokyo 113-0033, Japan
          \and Excellence Cluster ORIGINS, Boltzmannstraße, 2D-85748 Garching, Germany
          \and Univ. Grenoble Alpes, CNRS, IPAG, 38000 Grenoble, France
          \and Department of Astronomy \& Astrophysics, University of Toronto, 50 St. George Street, Toronto, ON M5S 3H4, Canada
          \and Instituto de Radioastronomía y Astrof\'isica, Universidad Nacional Autónoma de M\'exico, A.P. 3-72 (Xangari), 8701 Morelia, Mexico
          \and Instituto de Astronom\'ia, Univ. Nacional Autonoma de M\'exico, Ciudad Universitaria, A.P. 70-264, Ciudad de M\'exico 04510, Mexico
          \and RIKEN Cluster for Pioneering Research, 2-1, Hirosawa, Wako-shi, Saitama 351-0198, Japan
          \and SOKENDAI (The Graduate University for Advanced Studies), Shonan Village, Hayama, Kanagawa 240-0193, Japan
             }


 
  \abstract
   {Planets form in the disk surrounding young stars. When the planet formation process begins is still an open question. Annular substructures such as rings and gaps in disks are intertwined with planet formation, and thus their presence or absence is commonly used to investigate the onset of this process.}
   {Current observations show a limited number of disks surrounding protostars exhibiting annular substructures, all of them in the Class I stage. The lack of observed features in most of these sources may indicate a late emergence of substructures, but it could also be an artifact of these disks being optically thick. To mitigate the problem of optical depth, we investigate substructures within a very young Class 0 disk characterized by a low inclination using observations at longer wavelengths.}
   {We use 3 mm ALMA observations tracing dust emission at a resolution of 7 au to search for evidence of annular substructures in the disk around the deeply embedded Class 0 protostar Oph A SM1.}
   {The observations reveal a nearly face-on disk (inclination$\sim$16$^{\circ}$) extending up to 40 au. The  radial intensity profile shows a clear deviation from a smooth profile near 30 au, which we interpret as the presence of either a gap at 28 au  or a ring at 34 au with Gaussian widths of $\sigma=1.4^{+2.3}_{-1.2}$ au and $\sigma=3.9^{+2.0}_{-1.9}$ au, respectively. Crucially, the 3 mm emission at the location of the possible gap or ring is determined to be optically thin, precluding the possibility that this feature in the intensity profile is due to the emission being optically thick.}
   {Annular substructures resembling those in the more evolved Class I and II disks could indeed be present in the Class 0 stage, earlier than previous observations suggested. Similar observations of embedded disks in which the high optical depth problem can be mitigated are clearly needed to better constrain the onset of substructures in the embedded stages.  }


   \keywords{Stars: protostars -- Accretion, accretion disks -- protoplanetary disks --
                planets and satellites: formation --
                techniques: interferometric
               }

   \maketitle
%

\section{Introduction}

Recent observations have established that the planet formation process is already underway in $\gtrsim$ 1 Myr old Class II disks \citep{2018Andrewsdsharp,2018Manaramass,2021Benistypds70c,2022IzquierdoNew}. The finding of ubiquitous annular substructures (rings, gaps, cavities) in these disks at mm wavelengths with ALMA was critical to establish this \citep{2018HuangDisk,2018Longsubstructures}. This is because the annular substructures are either due to the presence of planets or they are places where dust grains can accumulate, a necessary step for grain growth that can lead to the formation of planetary cores \citep{2023Baesubstructuresppvii}. 

High-resolution observations resolving the dust emission in younger disks (Class 0/I with ages $\lesssim$ 0.5 Myr, \citealt{2014DunhamPPVI}) have been conducted to determine the emergence of annular substructures. Only a small number of these younger disks exhibit such substructures, and all of them are Class I \citep{2018SheehanGY91,2020SeguraCoxRing,2020NakataniSubstruc,2023OhashieDisk,2023YamatoL1489IRSedisk,2024ShoshiWL17,2024HsiehCAMPOSI}. This limited number of detections could imply that the planet formation process starts primarily during the later stages of protostellar evolution. Alternatively, the lack of substructures in most Class 0/I disks surveyed so far might be due to the observed emission being optically thick, particularly at 1.3 mm, which is the most common choice of wavelength. This is supported by multi-wavelength studies at a resolution from 70 to a few 100 au, which already suggested high optical depths at disk scales for embedded sources \citep{2007JorgensenPROSAC,2017Lialpha,2020KoResolving}. Recent multi-wavelength studies at higher-resolution ($\lesssim$15 au) also support high optical depths at 1.3 mm in young protostellar disks \citep{2018GalvanMadridselfobscuration,2021ApJLiuOMC3,2021ZamponiHotdisk,2022Maureirahotspots,2023WenruiTMC1A,2024Guerra-AlvaradoIRAS4A1}.

In this Letter, we use ALMA observations at a longer wavelength (3 mm) to resolve the disk continuum emission towards the deeply embedded Class 0 protostar Oph A SM1. Oph A SM1 (hereafter SM1) is located in the Oph A region in the Ophiuchus molecular cloud at a distance of 137 pc \citep{2018OrtizLeonGaia}. The spectral energy distribution (SED) is consistent with a cold $\sim 17$ K core with a mass of 1.3 \msun  \citep{2015PattkeOph}. The protostellar source is not detected at wavelengths $\leq$ 70 $\mu$m by Spitzer or Herschel/PACS \citep{2018FriesenALMA,2018KawabeDense}. Based on the upper limit to the flux at 70 $\mu$m, \cite{2018FriesenALMA} derived a maximum internal luminosity of $\sim$ 1 L$_{\odot}$\footnote{The relatively high upper limit is due to bright extended 70 $\mu$m emission near the location of SM1, see \cite{2018FriesenALMA} for details.}. No large scale outflow has been detected either. \cite{2018FriesenALMA} detected CO~(2-1) emission consistent with a low-velocity outflow ($< 6$ \kms) with a length $\lesssim$ 2000 au, implying a dynamical time of $1.6\times10^3$ yr  
 or less \citep{2018KawabeDense}. SM1 also shows variable hard X-ray emission, expected from young, actively accreting protostars \citep{2004GagneSimultaneous,2018KawabeDense}. Towards the compact source, a mass infall rate of $3\times10^{-5}$ \msun yr$^{-1}$ was derived from an inverse P Cygni profile detected in CO \citep{2018FriesenALMA}. Based on these observations, SM1 has been proposed as possibly the youngest protostar in Ophiuchus \citep{2014FriesenRevealing,2018FriesenALMA,2018KawabeDense}.

\section{Data}
\label{sec:data}
The ALMA band 3 observations using the C-10 configuration were taken in September 2021. The observations were part of the Cycle 7 project ID:2019.1.01074.S (PI: M. Maureira). The spectral setup consisted of four spectral windows with central frequencies between 93 and 105 GHz. The spectral setup was maximized for continuum observations, thus the maximum bandwidth of 1.875 GHz for each spectral window was used. These observations were planned to be combined with the C-6 configuration of the FAUST\footnote{Fifty AU Study of the chemistry in the disk/envelope system of Solar-like protostars} ALMA Large program \citep{2021CodellaFAUST} which covers the same frequency range (project ID:2018.1.01205.L, PI: S. Yamamoto). We used CASA \citep{2022CASATeam} to calibrate and image the data. We performed self-calibration separately for the observations in the extended and compact configuration, which were then combined and self-calibrated together. The cleaning in the self-calibration iterations was done first with the `multiscale' and later with the `mtmfs' deconvolver with nterms$=2$ for the final phase and amplitude steps. A robust parameter of 0 was set with 4 scales for cleaning in all steps. A manual mask was set and adjusted during the process when necessary. The details of the self-calibration iterations can be found in Appendix~\ref{ap:selfcal}. The self-calibration process improved the overall image fidelity and resulted in signal-to-noise S/N improvements of $\sim$40\% and $\sim$10\% for the extended configuration only, and combined dataset, respectively. The final synthesized beam and rms considering the region around SM1 are \SI{0.060}{\arcsecond}$\times$\SI{0.042}{\arcsecond} (at PA -79.5$^{\circ}$) and 25 $\mu$\jybm, respectively.

\section{Analysis}
\label{sec:analysis}

\begin{figure*}[h!]
   \centering
     \includegraphics[width=0.95\textwidth]{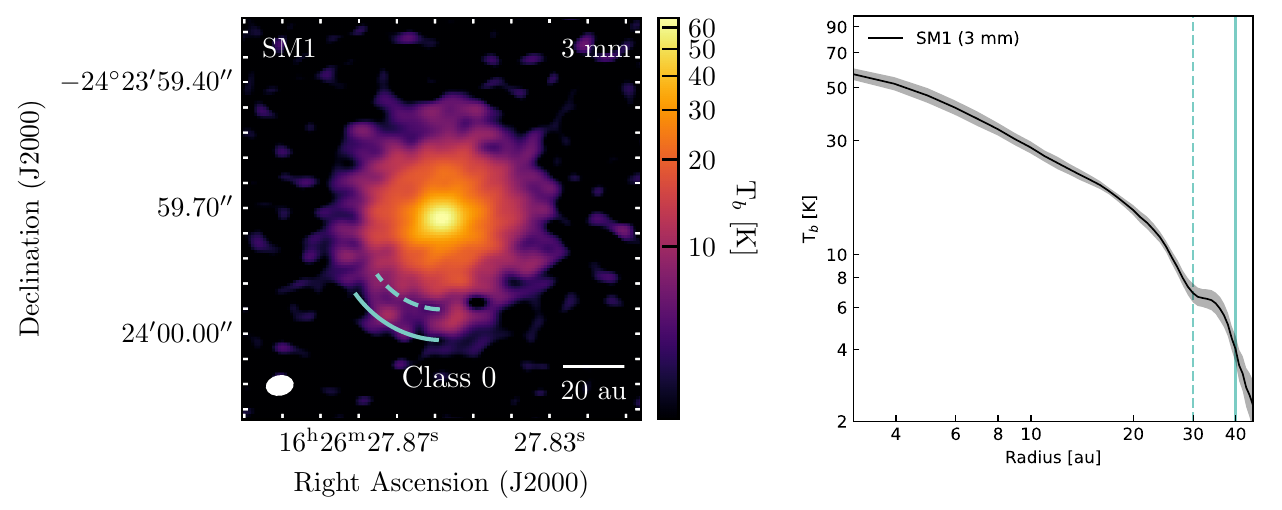}
      \caption{{\it Left: } 3 mm continuum observations of the SM1 disk. {\it Right:} Azimuthally averaged brightness temperature profile calculated using the full Planck function displayed in logarithmic scale. The corresponding $1\sigma$ uncertainties are shown in gray. Green dashed and solid vertical curves mark the same radial distance from the center in both panels, corresponding to the location of the deviation in the radial intensity profile and the location of the 3$\sigma$ emission contour in the map, respectively. }
       
         \label{fig:3mmobs_radprofile}
\end{figure*}

The left panel in Figure~\ref{fig:3mmobs_radprofile} shows the 3 mm continuum observations towards SM1. The observations reveal a disk structure with a nearly face-on orientation, with a brightness temperature peak of 57 K and a 3$\sigma$ contour extending up to $\sim$ 40 au from the center. Using CASA {\it imfit} we fit a 2D Gaussian to the emission resulting in center coordinates R.A.: 16:26:27.852, Dec.: -24:23:59.723, an inclination (assuming circular geometry) of $15.6^{\circ}$ and  P.A. of $147\pm35^{\circ}$. We used these values to deproject the emission, and produce an azimuthally averaged radial intensity profile\footnote{\url{https://github.com/jpinedaf/velocity_tools/blob/master/velocity_tools/coordinate_offsets.py}}. The right panel in Figure~\ref{fig:3mmobs_radprofile} shows the resultant intensity profile expressed in brightness temperature, calculated using the full Planck function. The error is calculated as $\sigma_{std}/\sqrt{N}$, where $\sigma_{std}$ and $N$ are the standard deviation and number of beams in each bin. 

Radial intensity profiles of disks are usually well described by a power-law in radius for the inner part, which is then truncated in the outer disk \citep{1998HartmannAcretion,2009IsellaStructure,2017TripathiSize,2021TazzariSizes}. For SM1, while the inner ($<20$ au) and the outermost ($> 40$ au) regions show such smoothly declining behavior there is a clear intensity deviation near 30 au. This type of feature is similarly observed in several intensity profiles from the DSHARP and ODISEA samples of Class II and Class I/II disks, respectively \citep{2018HuangDisk,2021MCiezaOdiseaIII}.

\subsection{Modelling the disk intensity profile}
\label{sec:fit}

We fit the full intensity profile with three different models: a smooth disk, a disk with a gap, and a disk with a ring. We first considered the intensity profile of a smooth disk. For this, we used a modified version of the self-similar profile \citep{1974LyndenBellEvolution} with distinct power-law indices constraining the inner and outer cutoff behaviors \citep{2019LongCompact,2019Manara,2021TazzariSizes}. The intensity profile has the form: 

\begin{equation}
    I(r) \propto  \left(\frac{r}{r_c} \right)^{\gamma_1}\exp\left[-\left(\frac{r}{r_c}\right)^{\gamma_2}\right]
\end{equation}

\noindent where $r$ is the radius, $r_c$ is the characteristic radius, $\gamma_1$ the inner disk power-law index, and $\gamma_2$ the power-law index describing the slope of the exponential cutoff. Together with $r_c$, $\gamma_1$ and $\gamma_2$, the fourth free parameter is taken as the total flux of the model, $F_{\rm tot}$. In contrast, the disk with a gap and the disk with a ring exhibit an intensity dip followed by an intensity peak annular features. While these structured disk models are visually similar, the difference comes from how these features are modeled. Following \cite{2019LongCompact}, we modeled these by considering either a smooth disk with a single carved gap or a smooth disk with a single ring exterior to the disk. Both the gap and ring are modeled as a Gaussian. The profile of the gap/ring annular substructure is given by $I_a\exp[(r-r_a)^2/(2\sigma_a^2)]$ where $r_a$, $I_a$, and $\sigma_a$ corresponding to the radius, intensity peak and width of the ring/gap respectively are additional free parameters. Each model was convolved with the synthesized beam and values for each of the parameters were obtained using the Python module \texttt{emcee} \citep{2013Foreman-Mackeyemcee}. Details of the fit setup and resultant values for the parameters as well as corner plots are in Appendix Section~\ref{sec:emcee_details}.

Figure~\ref{fig:fit_radprofile} shows, for all three models, the resultant fit and residuals. The smooth disk profile systematically overestimates and then underestimates the intensity near $\sim 30$ au where the deviation in intensity is observed. The middle and right panels show the results when considering a disk with an inner gap and an outer ring, respectively. Both models can reproduce the feature near $\sim30$ au and thus both lead to an improvement in the residuals. The models with gap and ring locate the annular substructure at a radius of $27.9^{+0.6}_{-0.6}$ au and $34.4^{+1.7}_{-2.1}$ au, respectively. The resultant $\sigma_a$ are $1.4^{+2.3}_{-1.2}$ au and $3.9^{+2.0}_{-1.9}$ au, for the gap and the ring, respectively. Table~\ref{table:fit_res} in the appendix summarizes the resultant values for the parameters in each model.

To compare the three models we use the Akaike Information Criterion (AIC)\footnote{\url{https://docs.astropy.org/en/stable/api/astropy.stats.akaike_info_criterion.html}}. 
The AIC is widely applied for model evaluation and selection \citep{2019CavanaughAIC,2021ChoudhuryTransition,2022ValdiviaMenaPeremb50}. It takes into account the goodness of the fit but also the possible over fitting when the number of free parameters increases as in the case of the non-smooth models considered here. The model with the lowest AIC is favored over the others. The AIC values are 37.8, 21.2, and 19.8 for the smooth disk, disk with a gap and disk with a ring, respectively. The smooth disk model can be excluded based on the significant difference\footnote{\url{https://docs.astropy.org/en/stable/api/astropy.stats.akaike_info_criterion.html}} ($\gtrsim10$) in AIC values compared with those including an annular substructure. On the other hand, although the AIC value is lower for the model with a ring the difference is not sufficient to favor one of the two non-smooth models. To summarize, non-smooth disk models are statistically preferred over a smooth disk model and the observed disk profile can be well reproduced by considering either a gap carved at 28 au within a disk with a radius\footnote{Radius containing 95\% of the flux.} of 40 au or a ring at 34 au outside a disk with a radius of 27 au.

\begin{figure*}[h!]
   \centering
     \includegraphics[width=0.95\textwidth]{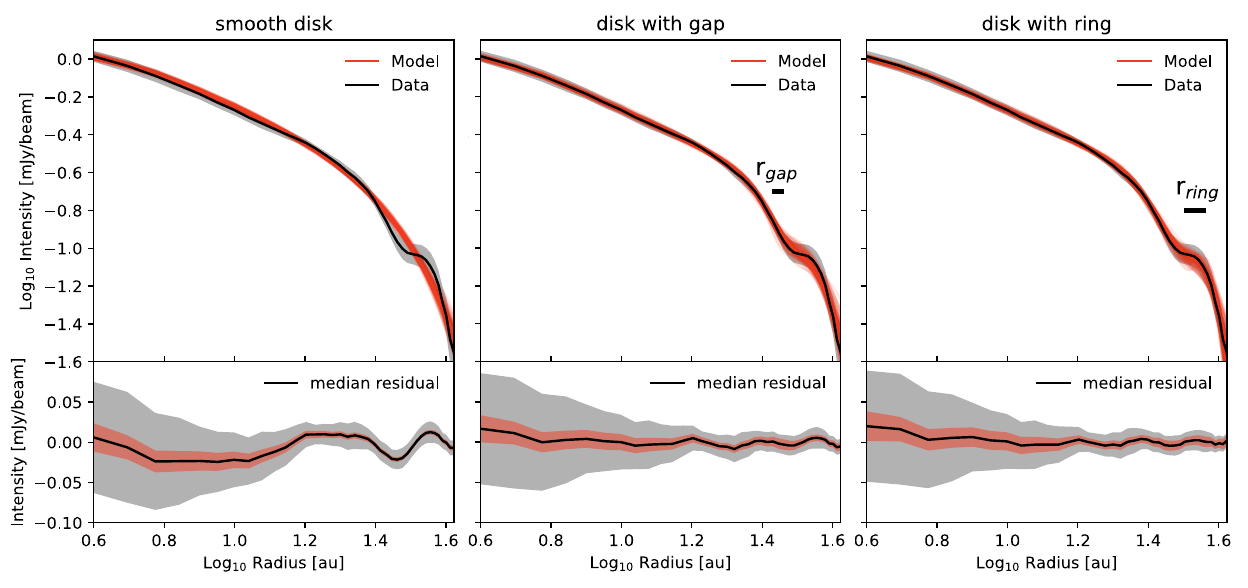}
      \caption{Fits to the disk intensity profile (top) and residuals (bottom). The orange curves in the top panels show 100 profiles, drawn randomly from the posterior distributions. The observed profile is shown with a black solid line.} The resultant median residual is shown in the bottom panels (solid black line) as well as 1$\sigma$ dispersion around the median residual (orange shaded area).  The gray shaded area corresponds to the $1\sigma$ uncertainty of the intensity profile in all panels. {\it Left: } Smooth disk profile. {\it Middle:} Disk profile with a gap at $\sim$ 28 au. The black horizontal segment shows the range of values for the gap radius enclosing 68\% of the posterior distribution around the median. {\it Right:} Disk profile with a ring at $\sim$ 34 au. The black horizontal segment shows the range of values for the ring radius enclosing 68\% of the posterior distribution around the median.
       
         \label{fig:fit_radprofile}
\end{figure*}

\subsection{Optical depth, temperature, and mass}
\label{sec:opt_depth}

Following \cite{2020SeguraCoxRing} and \cite{2016CiezaImaging}, we calculated the intra-band spectral index $\alpha$ by imaging the lower-sideband and upper-sideband spectral windows separately, using the parameters for imaging described in Section~\ref{sec:data}. We calculated the intensity profile for each image similar to Figure~\ref{fig:3mmobs_radprofile} and calculated alpha as  $\alpha = \ln(I_{\nu_1}/I_{\nu_2}) / \ln(\nu_1/\nu_2)$ where $\nu_1$ and $\nu_2$ correspond to 106.05 GHz and 94.05, respectively. Due to the limited sensitivity, the profile for $\alpha$ is limited to the inner $\sim$ 30 au. Beyond this radius $\alpha$ values are smaller than the $3\sigma$ statistical uncertainty, leaving $\alpha$ unconstrained. We note that besides the statistical uncertainties considered here, there is a systematic uncertainty at all radii of 0.1 due to a 0.8\% flux calibration error between spectral windows \citep{2020FrancisFluxCal}. 

\begin{figure}[h!]
   \centering
     \includegraphics[width=0.5\textwidth]{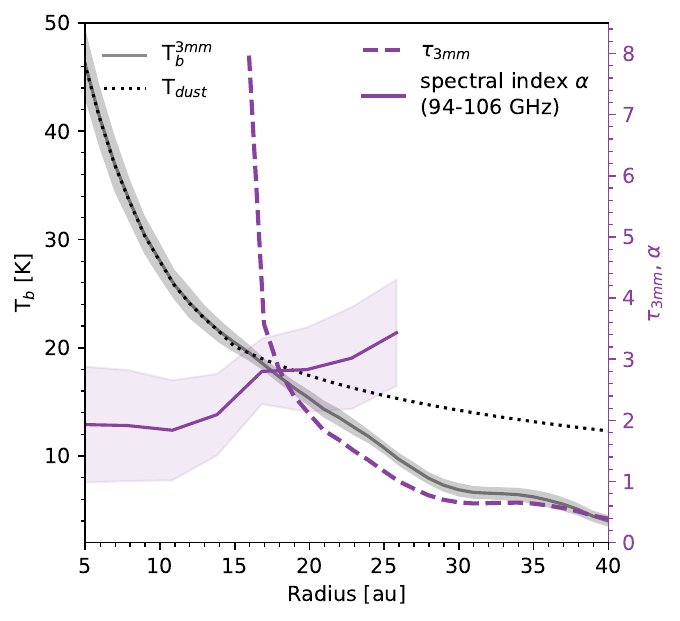}
      \caption{Radial profile of the band 3 (3 mm) in-band spectral index (solid purple). The shaded area shows the 1$\sigma$ uncertainty. The estimated 3 mm optical depth and temperature profiles are shown with dashed purple and dotted black lines, respectively. The temperature profile is extrapolated beyond r$\sim$15 au following a $r^{-0.5}$ power-law. The observed brightness temperature profile is overlaid in solid gray, similar to Figure~\ref{fig:3mmobs_radprofile}.}
       
         \label{fig:temp_alpha_tau}
\end{figure}

Figure~\ref{fig:temp_alpha_tau} shows the resultant spectral index profile overlaid with the intensity profile. At about 15 au the spectral index reaches the optically thick limit value of 2 and remains slightly below at small radii. Outside this radius it increases reaching values between 3 and 4, consistent with a reduction of the optical depth. To estimate the optical depth and dust temperature we followed \cite{2020SeguraCoxRing} and fit the emission of the profiles at 106 GHz and 94 GHz with a modified blackbody. In the inner region with the low spectral index values, the derived temperature converges to the brightness temperature. This is the expected behavior in the optically thick limit and thus it is also independent of the adopted value for the power-law index of the dust opacity $\beta$. At larger radii, temperature and optical depth are degenerate due to the emission becoming optically thinner. Following \cite{2020SeguraCoxRing}, we computed the temperature therein by extrapolating the temperature obtained in the optically thick regime ($r\lesssim15$ au) assuming a radial profile with a power-law exponent of $-0.5$ (passive heating). The resultant temperature and 3 mm optical depth profiles are overlaid in Figure~\ref{fig:temp_alpha_tau}. The temperature in the region with the gap/ring is about 15/13 K while the optical depth at 3 mm is below 1 therein. We note that we obtained a power-law exponent value $\gamma_1\sim-0.7$ from the intensity profile fit in the inner region of the disk (Table~\ref{table:fit_res}). This is steeper than expected one from passive heating and suggests the presence of accretion/viscuous heating. The presence of this type of non-radiative heating has been also suggested in other young disks (e.g., \citealt{2019LiuFuOri,2021ZamponiHotdisk,2022Maureirahotspots,2024TakakuwaeDisk}). Accretion/viscuous heating is expected to dominate in the inner parts of disks while irradiation is expected to still dominate the outer regions \citep{2015MordasiniGlobal}. Nevertheless, extrapolating the temperature profile outwards assuming such a steeper power-law exponent would further lower the temperature and therefore increase of the optical depth at a given radius 
by up to a 20\% and 30\%, respectively. This would still result in $\tau_{3mm}$ below one at the location of the gap/ring feature. 

The derived temperature is a factor of $\sim$2 lower than those derived at similar radii towards other disks surrounding embedded sources (hereafter embedded disks) with bolometric luminosities $\sim$ 1 $L_{\odot}$ and showing ring/gap structures \citep{2020SeguraCoxRing,2023SaieDiskC110IRS4}. The upper limit to the SM1 protostar luminosity is $\sim$1 L$_{\odot}$ \citep{2018FriesenALMA}. We therefore can estimate a lower limit to the source luminosity by assuming the difference in disk temperature between these sources is only due to luminosity (i.e., flaring is similar for these embedded disks and temperature in this region is set by irradiation only). In a passive disk the temperature and central source luminosity are related by $T_{mid}(r)\propto(\varphi L_*)^{1/4}/r^{1/2}$  where $\varphi$ is the flaring angle \citep{1997ChiangSpectral}. The 2$\times$ difference in temperature implies a 16$\times$ lower luminosity for SM1 of $\sim$0.08 $L_{\odot}$. Thus, it is possible that SM1 is a very low luminosity Class 0 protostar or VeLLO \citep{2008DunhamIndentifying}, in agreement with the very cold SED observed for this source. This would suggest that SM1 is either an extremely young protostar with very little mass yet accreted or a more evolved Class 0 protostar observed in a quiescent period of accretion \citep{2011PinedaL1451mm,2020MaureiraFHSC}. The former is favored by the observed outflow properties such as its short length and dynamical time \citep{2018FriesenALMA}. 

We also use the derived optical depth profile $\tau_{3mm}$ in Figure~\ref{fig:temp_alpha_tau} to estimate the mass of the disk as $M_{disk} = 2\pi \int_0^{R} \Sigma(r) r dr$
where $\Sigma(r) =\tau_{3mm}(r)/\kappa_{3mm}$ and $\kappa_{3mm}$ is the dust opacity. We used a constant value for $\tau_{3mm}$ of 8 for the inner $\sim$ 15 au region (see Figure~\ref{fig:temp_alpha_tau}) and $R=40$ au to obtain a conservative estimation. Using an opacity at 3 mm of $\sim$ 1 cm$^2$ gr$^{-1}$ \citep{1990beckwithSurvey, 2018Birnsteil} and a gas to dust ratio of 100 we estimate a total gas mass of 0.13 \msun (or $\sim$ 430 M$_{\oplus}$ dust mass). About half of this mass correspond to the outer region with $\tau_{3mm}<8$. The estimated mass in solids for this disk is in agreement with the range derived for a sample of Class 0 disks also using long wavelength observations (9 mm), resulting in masses 50$\times$ higher than those of Class II disks \citep{2020TychoniecDust}. Nevertheless, we note that the derived mass can be higher or lower in a factor of 2-5 due to uncertainties on the dust opacities which can show large variations depending on the considered grain properties (e.g., \citealt{2010RicciDust,2018Birnsteil}).

\section{Discussion}
\label{sec:discussion}

In this work, we show that both a disk with a gap and a disk with an outer ring can explain the observed deviation in the intensity profile of this young Class 0 disk. This type of deviation or inflection point in the intensity profile has also been interpreted as unresolved gaps/rings for several Class II disks, and interestingly, in those cases it also appears to be located near the edge of the disk structure (e.g., WSB 52, Sz 129 and RU Lup in \citealt{2018HuangDisk}). Similarly, the Class I disk IRS 63 was revealed to have a ring and a gap near the location of two inflection points in its intensity profile \citep{2020SeguraCoxRing,2021MCiezaOdiseaIII}. The lifetime of Class 0 protostars, such as SM1, is estimated to be $\sim0.15$ Myr \citep{2014DunhamPPVI}, therefore the evidence for a gap or a ring in SM1 suggests a fast emergence of dust substructures in disks. This is in agreement with several observational studies showing that the amount of solid material in Class II disks only about 1 Myr old is not enough to explain the observed exoplanets systems \citep{2018Manaramass,2020TychoniecDust,2022TestiPPpopulation}. In addition, the evolution of the dust mass for Class II disks can be explained by internal regeneration of dust due to early ($\lesssim$0.5 Myr) planet formation \citep{2022TestiPPpopulation,2022BernaboDustResurgence}. Thus, substructures at early stages as the one found here could indeed indicate the presence of a planet, or promote a fast subsequent formation of a planetary core. Theoretically, the possibility of a planet at such early times depends on whether the conditions for the streaming instability (dust-to-gas ratio near unity and dust grains only moderately coupled to the gas), could be present in such a young disk \citep{2005YoudinSI}. In recent years, theoretical studies have started to investigate this question. \cite{2022CridlandEarlyPlanet} find that the conditions for streaming instability are met on timescales of 0.1-0.2 Myr for regions in the disk at a distance of few tens of au, with masses from few M$_{\oplus}$ to tens M$_{\oplus}$ available to form planetesimals. The favorable conditions are aided by the early decoupling of grains with sizes $\gtrsim$ 10 $\mu$m in the infalling envelope and disk \citep{2017BateDustDynamics,2020LebreuillyDustRichDisks,2022CridlandEarlyPlanet}. Such masses agree with the possible range of planetary masses derived from the gap properties in this work although there are large uncertainties (see Appendix Section~\ref{ap:planetmass}). On the other hand, other studies find that the conditions are not met in young Class 0/I disks as their high turbulence and density would prevent the growth, decoupling and settling of dust (e.g., \citealt{2018DrazkoskaDiskBuildup,2023XuRevisitingDustGrowth}). In the more conservative scenario in which an early substructure, such as the one in SM1, precedes the formation of planetesimals, numerical simulations find that in young embedded disks, infalling dust can accumulate at the outer disk's edge resulting in a ring structure \citep{2022KuznetzovaAnisotropic}. This scenario therefore aligns with the disk plus ring model, found in Section 3, as a plausible explanation for the observed deviations in the SM1 disk intensity profile. Moreover, recent numerical simulations show that such dust accumulation could start even before the formation of the protostar \citep{2024BhandareMixing}. In summary, the discussion above suggests that although several independent observations, including our work, support the early formation of planets during the embedded Class 0/I stages, theory studies are yet to converge on whether planetesimal formation mechanisms can be effective in these disks.\\

Besides the above discussed possibilities, other scenarios that could potentially lead to the observed deviation in the intensity profile of SM1 are changes in the dust opacity at the location of icelines \citep{2016CiezaImaging} or due to the variation of grain sizes, as well as dust accumulation at pressure bumps caused by planets \citep{2021MCiezaOdiseaIII,2019PinillaInnerDisk}. The derived temperature at the gap/ring location  ($\sim15/13$ K) is close to freeze-out temperature range of N$_2$ (17-21 K, \citealt{2016FayolleN2,2018HuangDisk}), while the CO and CO$_2$ icelines fall in the optically thick part of the disk. Thus, it is possible that the deviation is due to the location of the N$_2$ iceline, which can lead to changes in the dust properties. Multi-wavelength and high-resolution observations are needed to better constrain the temperature and grain properties at this location (e.g., \citealt{2021MaciasTW,2024Guerra-AlvaradoBand9HLTau}). \\

Finally, when the emission from the disk becomes optically thick the intensity starts to trace the temperature profile instead of a combination of the temperature and optical depth \citep{2024Guerra-AlvaradoBand9HLTau}. This makes the detection of substructures more difficult, and thus can result in an underestimation of the frequency of substructure in the younger ($\lesssim0.5$ Myr) embedded disks \citep{2023OhashieDisk,2024HsiehCAMPOSI}. The problem is exacerbated if the disk is observed at higher inclinations, adding uncertainty on the origin of features in the profile (e.g., \citealt{2023SaieDiskC110IRS4}). This may help explain the low frequency of substructure found thus far in embedded disk, and will be discussed in more detail in a forthcoming paper. We predict that the optical depth at 1.3 mm for SM1 would indeed be above one within 40 au of the central source which, depending on the assumed dust properties, can indeed reduce the contrast of the observed deviation in the profile (see Appendix Figure~\ref{fig:tau_pred}) making its detection more challenging. Our observations at longer wavelengths, combined with the low inclination of the source, have resulted in optically thin emission beyond 20 au (Figure~\ref{fig:temp_alpha_tau}), making SM1 a more robust case for the detection of substructure in a deeply embedded Class 0 source. 

\section{Summary and conclusions}
\label{sec:sum_conclusion}
We use ALMA observations at 3 mm with a resolution of 6.9 au to probe substructures in the deeply embedded Class 0 disk surrounding Oph A SM1. The observations reveal a nearly face-on disk with a radius of $\sim$ 40 au. The radial intensity profile shows deviation near 30 au that can be reproduced either by considering a gap at radius of 28 au or an outer ring at a radius of 30 au. A smooth profile is statistically excluded. The 3 mm emission at the location of the substructure is optically thin, which is a critical condition to reveal such features. The evidence from SM1 suggests that we should be cautious that annular substructures might emerge early, during the Class 0 stage ($\lesssim0.2$ Myr), but remain difficult to identify in current observations. More observations using
longer wavelengths, or targeting disks in which the optical depth
can be mitigated are needed to more robustly constrain the onset and frequency
of substructures during the embedded stages. 
  
\begin{acknowledgements}
      The authors thank Masayuki Yamaguchi and Munan Gong for insightful discussions on imaging techniques and disk theory, respectively. The authors also kindly thank the anonymous referee for the comments that helped improve the manuscript. Part of this work was supported by the Max-Planck Society. D.M.S. is supported by an NSF Astronomy and Astrophysics Postdoctoral Fellowship under award AST-2102405. ClCo, LP, and GS acknowledge the PRIN-MUR 2020 BEYOND-2p (``Astrochemistry beyond the second period elements'', Prot. 2020AFB3FX), the project ASI-Astrobiologia 2023 MIGLIORA (Modeling Chemical Complexity, F83C23000800005), the INAF-GO 2023 fundings PROTO-SKA (Exploiting ALMA data to study planet forming disks: preparing the advent of SKA, C13C23000770005), the INAF Mini-Grant 2022 “Chemical Origins” (PI: L. Podio), INAF-Minigrant 2023 TRIESTE (``TRacing the chemIcal hEritage of our originS: from proTostars to planEts''; PI: G. Sabatini), and the National Recovery and Resilience Plan (NRRP), Mission 4, Component 2, Investment 1.1, Call for tender No. 104 published on 2.2.2022 by the Italian Ministry of University and Research (MUR), funded by the European Union – NextGenerationEU– Project Title 2022JC2Y93 Chemical Origins: linking the fossil composition of the Solar System with the chemistry of protoplanetary disks – CUP J53D23001600006 - Grant Assignment Decree No. 962 adopted on 30.06.2023 by the Italian Ministry of Ministry of University and Research (MUR). 
      L.L. acknowledges the support of DGAPA PAPIIT grants IN108324 and IN112820 and CONACyT-CF grant 263356. This project has received funding from the European Research Council (ERC) under the European Union Horizon Europe programme (grant agreement No. 101042275, project Stellar-MADE). Y.A. acknowledges support by NAOJ ALMA Scientific Research Grant code 2019-13B.
\end{acknowledgements}

%
%

\bibliographystyle{aa} 
\bibliography{SM1_disk.bib} 

\begin{appendix}
\section{Data self-calibration}
\label{ap:selfcal}
For the extended configuration observations (C-10) we performed phase self-calibration only. We performed several iterations reducing the solint interval from 'inf' down to '10s', combininig all the spectral windows for the final step. This phase-only self-calibration process improved the S/N of the image by about 40\%. For the more compact (C-6) configuration, there were seven spectral windows, two of which contained spectral line emission. Line-free channels were used to obtain a continuum model that was then used for self-calibration using solint$=$'int' (per 6s integration), while per-scan (8-10 minutes) solutions were used for amplitude and phase self-calibration once the model was sufficiently complete. For the combined self-calibration, we averaged both datasets in time and frequency. We averaged in time in bins of 9 seconds. We averaged in frequency up to the maximum that would not result in a reduction of the intensity larger than $\sim$1\% for sources away from the phase center due to bandwidth
smearing\footnote{\url{https://safe.nrao.edu/wiki/pub/Main/RadioTutorial/BandwidthSmearing.pdf}}. The primary beam response at the location of SM1 is 0.35 because the target of this field was the Class 0 source VLA 1623-2417. The final channel width used was 30 MHz.
We performed four successive steps of phase-only self-calibration in the combined dataset. In the first one, we use solint$=$'inf', gaintype$=$'G' and combined all spectral windows. For this first round, we used the final image after self-calibration of the extended configuration as the model. This aligns the two datasets to the peak position of the extended configuration and thus corrects for differences in the peak positions across datasets arising from proper motion or astrometric errors of the observations. The resultant model was then used for the next steps all with gaintype$=$'T'. The solint values for the next iterations were 'inf', '30s', 'int'. For the last two iterations all spectral windows were combined. Finally, we performed three steps of amplitude self-calibration with solint$=$'inf'. In the first one we combined all spectral windows and used solnorm$=$T, then we switched to solnorm$=$T and finally we allowed solutions to be calculated per spectral window. For the combined self-calibration the improvement in S/N was about 10\%.

\section{Fit of the intensity profiles}
\label{sec:emcee_details}

\begin{figure*}[h!]
   \centering
     \includegraphics[width=0.9\textwidth]{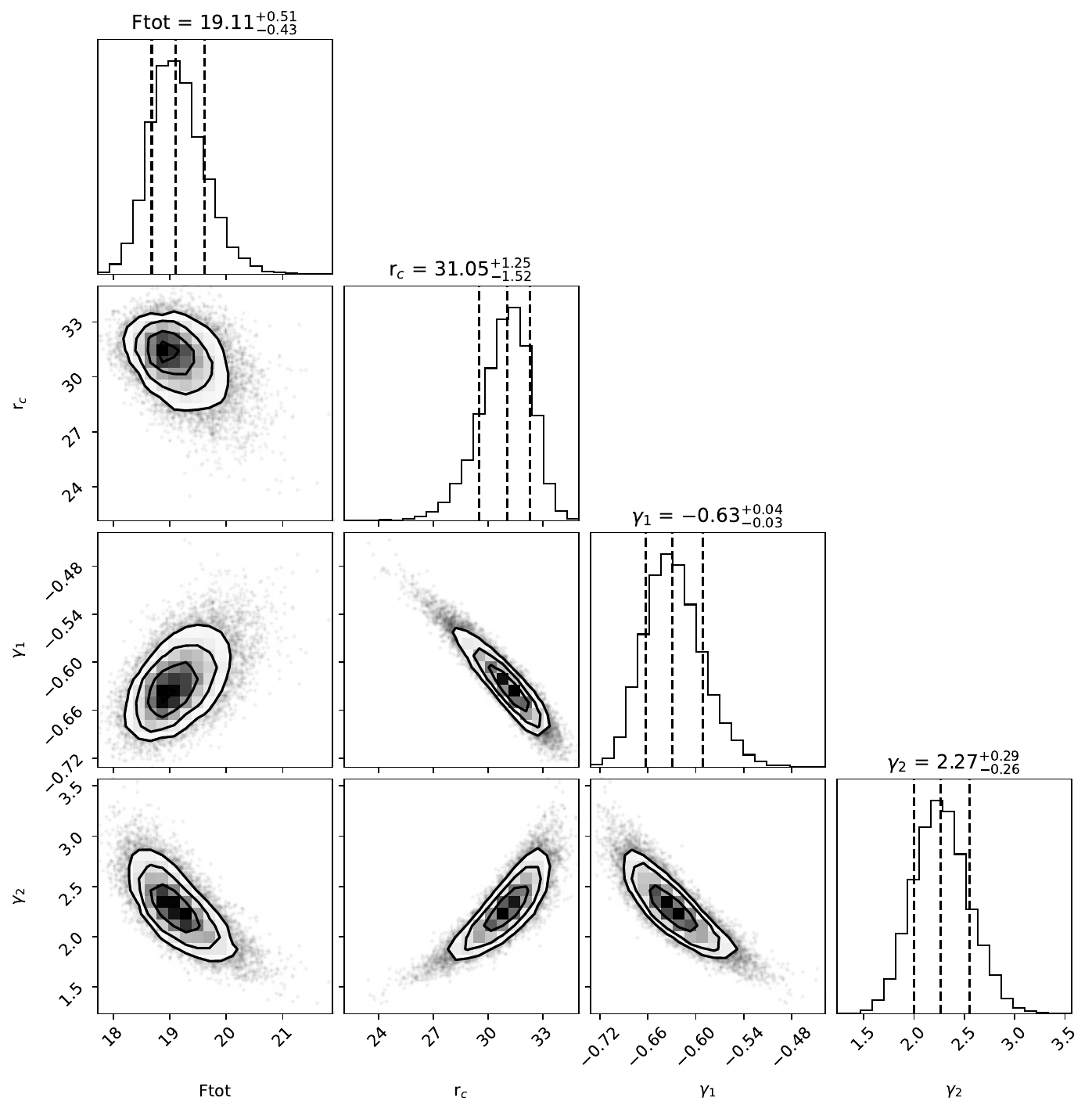}
      \caption{Corner plot results for the smooth disk profile. The dashed lines marked the median value and the lower and upper range enclosing the 68\% of the distribution around the median.}
       
         \label{fig:corner_smooth}
\end{figure*}

\begin{figure*}[h!]
   \centering
     \includegraphics[width=1.0\textwidth]{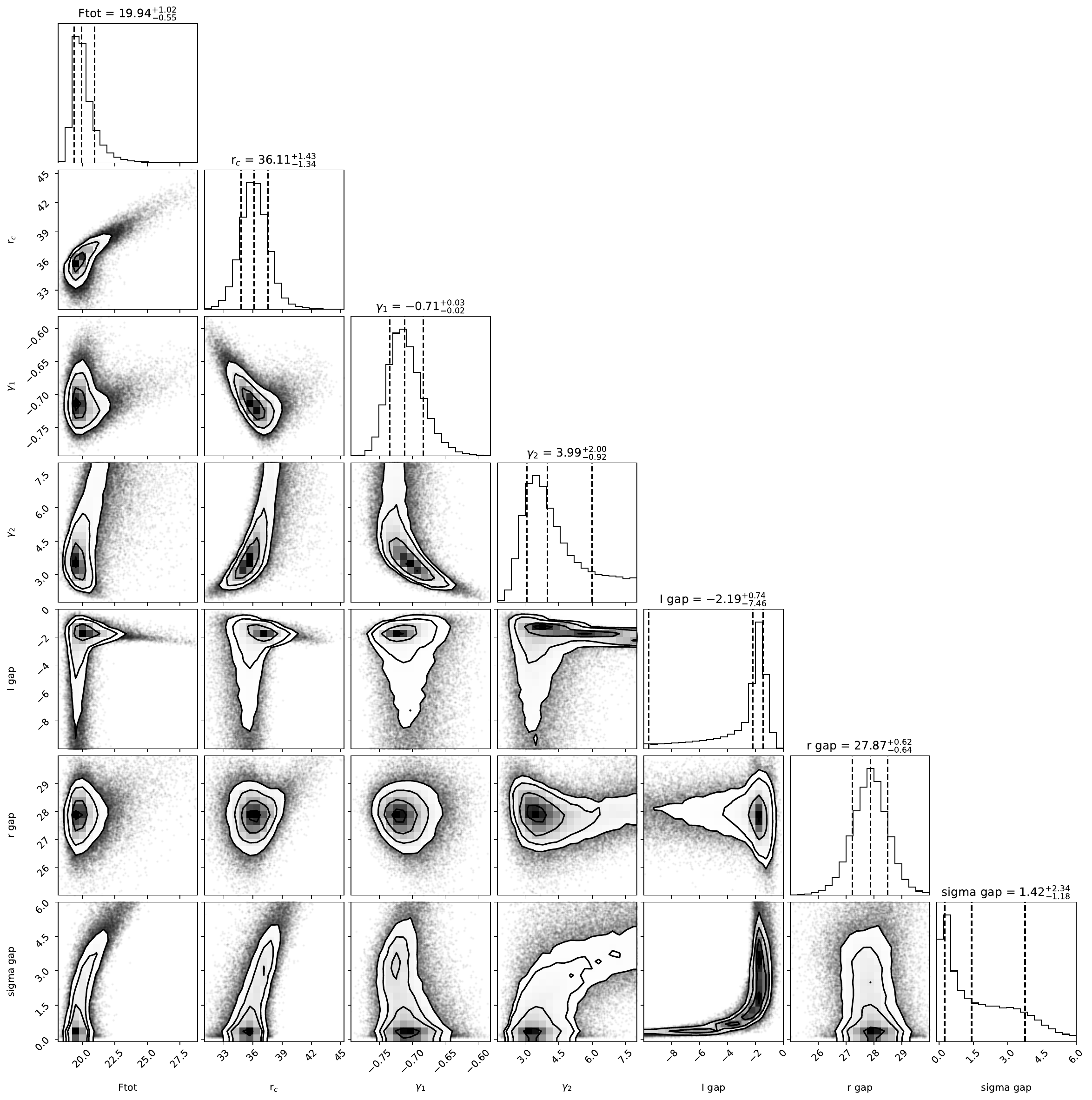}
      \caption{Similar to Figure~\ref{fig:corner_smooth} for the model of a disk with a gap.}
       
         \label{fig:corner_gap}
\end{figure*}

\begin{figure*}[h!]
   \centering
     \includegraphics[width=1.0\textwidth]{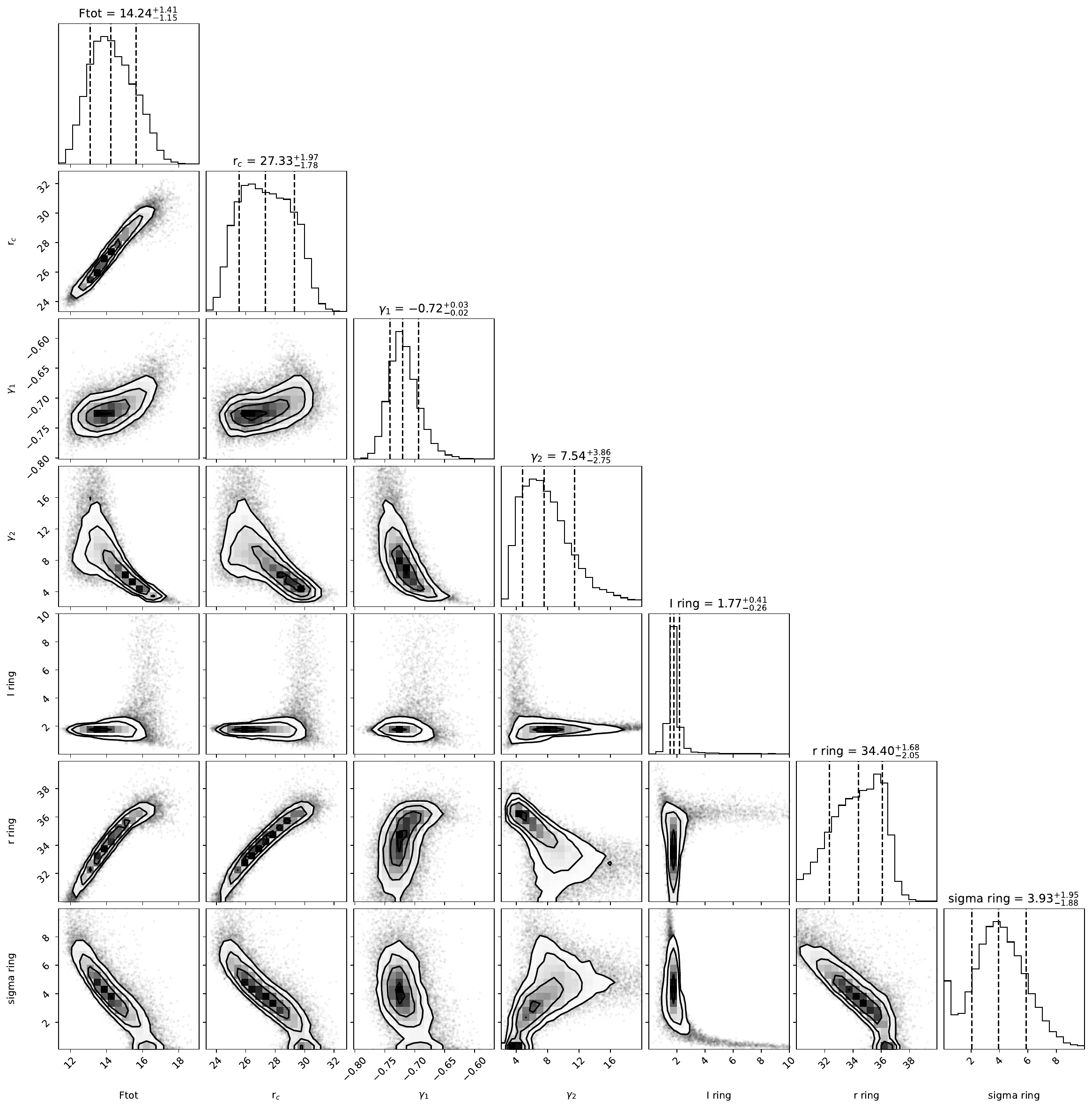}
      \caption{Similar to Figure~\ref{fig:corner_smooth} for the model of a disk with a ring.}
       
         \label{fig:corner_ring}
\end{figure*}

We use \texttt{emcee} \citep{2013Foreman-Mackeyemcee} to fit the beam convolved models to the data. The logarithms of the convolved model and the data were used in the likelihood function in order to consider the fractional errors instead of the absolute errors. This is to avoid a higher weight on the brighter parts of the disk with respect to the weaker extended outer parts. The convolution of the model was done using the function \texttt{fftconvolve} from the Python module \texttt{scipy.fftconvolve} \citep{2020SciPyVirtanen}. We used uniform priors for each parameter which are summarized for the individual models in Table~\ref{table:fit_priors}. We use 50 walkers and run it up to a maximum of 300,000 steps stopping the run when the number of steps was larger than 100$\times$ the autocorrelation time (which is the effective number of independent steps) and the autocorrelation time estimation varied by less than 1\%. The burn-in was computed as twice the autocorrelation time and the chains were thinned using a thinning parameter of half the autocorrelation time. The resulting chains showed no large-scale variations. Figures~\ref{fig:corner_smooth} to~\ref{fig:corner_ring} show the resultant corner plots. Table~\ref{table:fit_res} summarized the resultant median for the free parameters in each model with their respective lower and upper range enclosing 68\% of the distribution around the median. The last row shows the radius of the disk enclosing the 95\% of the total flux $R_{95}$. We note that for the gap model there is a degeneracy between models with a deep and narrow gap or a shallow broad one as both lead to a similar profile once convolved with the beam. Given that the posteriors of the intensity peak and width for the gap models show significant tension, extreme values for the width (small or large) seem less likely. Higher resolution observations are required to improve the constrains on the gap morphology.

\begin{table*}
\caption{Results from the fit. The equations used for the models are described in Section~\ref{sec:fit}. $r_c$ is the characteristic disk radius, $\gamma_1$ is the inner disk power-law index and $\gamma_2$ is the power-law index describing the slope of the exponential cutoff,  and $F_{\rm tot}$ is total disk flux. The parameters describing the Gaussian annular substructure are  $r_a$, $I_a$, and $\sigma_a$, corresponding to the radius, intensity peak and width of the ring/gap respectively.}
\label{table:fit_res}      
 \centering
 \renewcommand{\arraystretch}{1.2}                                                                                  
\begin{tabular}{l r r r }
\hline\hline
Parameter & smooth disk & disk with a gap & disk with a ring \\    
\hline                        %
F$_{tot}$ [mJy] &	$19.11^{+0.51}_{-0.43}$ & $19.94^{+1.02}_{-0.55}$ & $14.24^{+1.41}_{-1.15}$ \\
r$_c$ [au] &  $31.1^{+1.3}_{-1.5}$  & $36.1^{+1.4}_{-1.3}$  & $27.3^{+2.0}_{-1.8}$ \\
$\gamma_1$ & $-0.63^{+0.04}_{-0.03}$ & $-0.71^{+0.03}_{-0.02}$ & $-0.72^{+0.03}_{-0.02}$\\
$\gamma_2$ & $2.27^{+0.29}_{-0.26}$ & $3.99^{+2.00}_{-0.92}$ & $7.54^{+3.86}_{-2.75}$\\
$I_a$ [$\mu$Jy au$^{-2}$]& ... & $-2.19^{+0.74}_{-7.46}$ & $1.77^{+0.41}_{-0.26 }$ \\
$r_a$ [au] & ... &$27.9^{+0.6}_{-0.6}$  & $34.4^{+1.7}_{-2.1}$ \\
$\sigma_a$ [au] & ... &$1.4^{+2.3}_{-1.2}$  & $3.9^{+2.0}_{-1.9}$ \\
\hline
$R_{95}$ [au]& $43.5^{+2.5}_{-2.5}$ &  $39.5^{+2.5}_{-2.5}$& $26.5^{+4.0}_{-2.0}$\\

\hline                        %

\end{tabular}
\end{table*}

\begin{table*}
\caption{Priors for fit. The parameters are the same as in Table~\ref{table:fit_res}.}
\label{table:fit_priors}      
 \centering
 \renewcommand{\arraystretch}{1.2}                                                                                  
\begin{tabular}{l r r r }
\hline\hline
Parameter & smooth disk & disk with a gap & disk with a ring \\    
\hline                        %
F$_{tot}$ [mJy] &	$(0,30)$ & $(18,29)$ &  $(10,29)$ \\
r$_c$ [au] & $ (10,70)$ & $[31,47)$  &  $[10,33]$\\
$\gamma_1$ & $(-10,1)$ & $(-1,0.01)$ &  $(-1,-0.01)$\\
$\gamma_2$ & $(0,20)$ & $(1,8)$ &  $(1,20)$\\
$I_a$ [$\mu$Jy au$^{-2}$]& ... & $(-100,-0.1)$ &  $(0.1,100)$\\
$r_a$ [au] & ... &$(25,30]$  &  $(30,40]$ \\
$\sigma_a$ [au] & ... &$(0.1,10)$&  $(0.1,10)$\\

\hline                        %

\end{tabular}
\end{table*}

\section{Intensity and optical depth predictions at 1.3 mm}

\begin{figure}[h!]
   \centering
     \includegraphics[width=0.9\hsize]{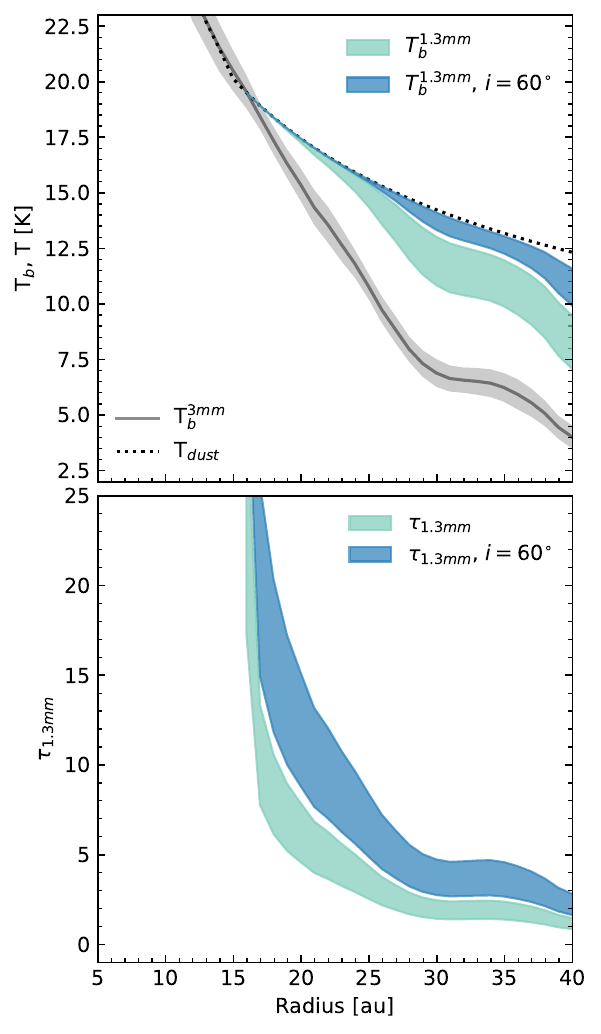}
      \caption{Predictions for the brightness temperature and optical depth profiles at 1.3 mm based on the observed 3 mm profile for SM1 (solid gray line). The widths of the shaded areas correspond to the resultant range of the profiles, considering $\beta$ between 1 and 1.7. The darker blue corresponds to the predicted profile if the disk had been observed at a higher inclination ($i=60^{\circ}$). The black dotted line corresponds to the derived temperature for SM1 as in Figure~\ref{fig:temp_alpha_tau}.}
       
         \label{fig:tau_pred}
\end{figure}

Figure~\ref{fig:tau_pred} shows the predicted 1.3 mm intensity profiles and optical depth for SM1 by extrapolating $\tau_{3mm}$, using constant $\beta$ values between 1 and 1.7, and the temperature profile derived in Section~\ref{sec:opt_depth}. At 1.3 mm the optical depth increases reaching values greater than 1 throughout the profile. The increase in optical depth also reduces the contrast of the gap/ring feature as the intensity approaches the temperature profile. Similarly, the increased optical depth leads to a flattening or `bump' in the profile from 15 to 25 au. This feature in the profile, which is interior to the observed 3 mm intensity feature, and therefore not due to the presence of dust substructure but due to the optically thick emission. Figure~\ref{fig:tau_pred} also shows the predictions for SM1 if it were inclined by 60$^{\circ}$. This inclination corresponds to an increase in optical depth by a factor of about 2. The predictions show that in such situations the profile could appear featureless (upper edge dark blue shaded region). 

\section{Estimation of planetary mass from gap properties}
\label{ap:planetmass}

Numerical simulations show that when a gap is opened due to the presence of a planet, the width of the gap normalized by its distance to the star is correlated with the ratio between the star and the planet masses. The exact relation also depends on the thickness of the disk (H/R) at the location of the planet and on the viscosity parameter $\alpha_{vis}$ \citep{2016KanagawaMassConstraint,2016RosottiMinimum}. In the case of SM1, several of these parameters are unknown, including the protostar mass. Following Equation 4 in \cite{2016KanagawaMassConstraint} and assuming $\alpha_{vis}=10^{-3}$, an aspect ratio H/R$\sim$0.1-0.3 (following the \citealt{2020SeguraCoxRing} derivation for the Class I disk IRS 63) and a 1 M$_{\odot}$ protostar, we obtain possible planet masses in the range from few tens to 200 M$_{\oplus}$. The planet's mass scales linearly with the protostar mass. A comparable range is found when compared with the numerical simulations by \cite{2016RosottiMinimum}. We should be cautious not to overinterpret these values given the number of unknown parameters. Nonetheless, the potential upper range agrees with the estimated planet masses for the gaps in the Class I disk IRS 63, and the possible range agrees with numerical estimates of the mass budget for planetary core formation through the streaming stability in young disks \citep{2022CridlandEarlyPlanet}. Molecular line observations are required to obtain the protostar mass for SM1, which would allow us to set tighter constraints on both the disk's dynamical state and the possible planetary masses.

\end{appendix}

\end{document}